\documentclass[11pt]{article}
\usepackage{epsfig}
\graphicspath{{/}}

\begin{document}
\baselineskip 24pt
\begin{center}
{\large\bf Electron-Hole Generation and Recombination Rates for Coulomb Scattering in Graphene} \\
\normalsize
\vspace{5em}
Farhan Rana$^{1}$ \\	
\vspace*{2em}
{\em $^{1}$School of Electrical and Computer Engineering, Cornell University, Ithaca, NY 14853 \\}
\vspace{4em}
\large\bf Abstract \\
\normalsize
\end{center}
We calculate electron-hole generation and recombination rates for Coulomb scattering (Auger recombination and impact ionization) in Graphene. The conduction and valence band dispersion relation in Graphene together with energy and momentum conservation requirements restrict the phase space for Coulomb scattering so that electron-hole recombination times can be much longer than 1 ps for electron-hole densities smaller than $10^{12}$ cm$^{-2}$.  
\newpage 

\section{Introduction}
Graphene is a single two dimensional (2D) atomic layer of carbon atoms forming a dense honeycomb crystal lattice~\cite{dressel}. The electronic properties of Graphene have generated tremendous interest in both experimental and theoretical arenas~\cite{chakra,ryzhii,nov1,nov2,zhang,heer}. The massless energy dispersion relation of electrons and holes with zero (or close to zero) bandgap results in novel behavior of both single-particle and collective excitations~\cite{dressel,chakra,ryzhii,nov1,nov2,zhang,heer}. The high mobility of electrons in Graphene has prompted theoretical and experimental investigations into Graphene based ultra high speed electronic devices such as field-effect transistors, pn-junction diodes, and terahertz oscillators~\cite{ryzhii,nov2,lundstrom,marcus,rana,ryzhii2,ryzhii3}. The behavior of many of these devices depends on the electron-hole recombination rates in Graphene. For example, the diffusion length of injected minority carriers in a pn-junction diode is proportional to the square-root of the minority carrier recombination time~\cite{pierret}. It is therefore important to understand the mechanisms that are responsible for electron-hole generation and recombination in Graphene and the associated time scales. 

Small band-gap semiconductors usually have large electron-hole recombination rates due to Coulomb scattering (Auger recombination)~\cite{paul}. Graphene, with a zero (or close to zero) bandgap, presents a limiting case. The zero bandgap and the large optical phonon energy in Graphene (196 meV~\cite{ando}) suggest that electron-hole recombination rates could be dominated by Auger processes. In addition, the zero bandgap also implies that electron-hole generation rates in Graphene due to Coulomb scattering (impact ionization) may also not be small even in the absence of high energy carriers.    

In this paper we calculate the electron-hole generation and recombination rates for Coulomb scattering (Auger recombination and impact ionization) in Graphene. We show that the conduction and valence band dispersion relation in Graphene together with energy and momentum conservation requirements restrict the phase space for Coulomb scattering so that electron-hole generation recombination times can be much longer than 1 ps at all temperatures for electron-hole densities smaller than $10^{12}$ cm$^{-2}$.      

\begin{figure}[tp]
  \begin{center}
   \epsfig{file=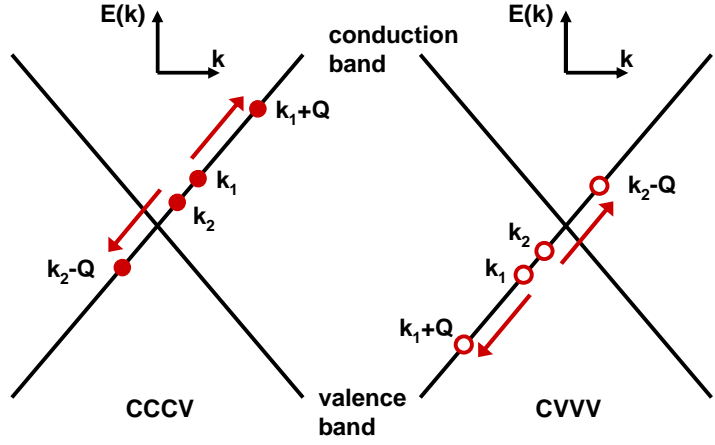,angle=-90,width=4.0 in}    
    \caption{Electron-hole recombination in Graphene from Coulomb scattering (Auger recombination) via the CCCV and the CVVV processes is shown. The two processes shown are mirror images of each other.}
    \label{Fig1}
  \end{center}
\end{figure}
\begin{figure}[bp]
  \begin{center}
   \epsfig{file=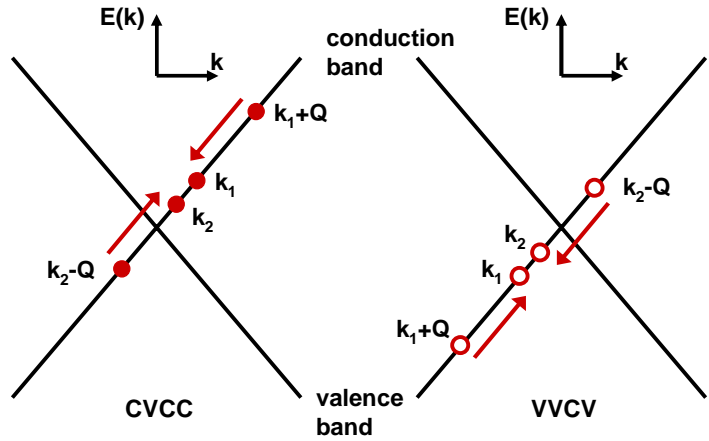,angle=-90,width=4.0 in}    
    \caption{Electron-hole generation in Graphene from Coulomb scattering (impact ionization) via the CVCC and the VVCV processes is shown. The two processes shown are mirror images of each other. In addition, the generation processes shown are the reverse of the recombination processes shown in Fig.1.}
    \label{Fig2}
  \end{center}
\end{figure}

\section{Theoretical Model}
In Graphene, the valence and conduction bands resulting from the mixing of the $p_{z}$-orbitals are degenerate at the inequivalent $K$ and $K'$ points of the Brillouin zone~\cite{dressel}. Near these points, the conduction and valence band dispersion relations can be written compactly as,
\begin{equation}
E_{s}({\bf k}) = s \hbar v |{\bf k}|
\end{equation}
where $s=\pm 1$ stand for conduction ($+1$) and valence ($-1$) bands, respectively, and $v$ is the ``light'' velocity of the massless electrons and holes, and assumed to be equal to $10^{8}$ cm/s. The wavevector ${\bf k}$ is measured from the $K$($K'$) point. 

Electron-hole recombination due to Coulomb scattering (Auger recombination) in Graphene occurs by the two processes depicted in Fig.1. In the CCCV process (Fig.1), an electron in the conduction band with initial momentum ${\bf k}_{1}$ scatters off another electron in the conduction band with momentum ${\bf k}_{2}$. The result is an electron in the conduction band with momentum ${\bf k}_{1}+{\bf Q}$ and an electron in the valence band with momentum ${\bf k}_{2}-{\bf Q}$. In the CVVV process (Fig.1), a hole in the valence band with initial momentum ${\bf k}_{1}$ scatters off another hole in the valence band with momentum ${\bf k}_{2}$. The result is a hole in the valence band with momentum ${\bf k}_{1}+{\bf Q}$ and a hole in the conduction band with momentum ${\bf k}_{2}-{\bf Q}$. The CVVV process is a mirror image of the CCCV process. The labeling scheme used here indicates the initial and final states of the two electrons involved in the scattering process. The electron-hole recombination rates $R_{\rm CCCV}(n,p)$ and  $R_{\rm CVVV}(n,p)$ are functions of the electron and hole densities, $n$ and $p$, respectively, and the symmetry between the conduction and valence band dispersions in Graphene implies that $R_{\rm CCCV}(n,p) = R_{\rm CVVV}(p,n)$. Electron-hole generation due to Coulomb scattering (impact ionization) occurs by the reverse of the CCCV and CVVV recombination processes. We label these generation processes as CVCC and VVCV, respectively, and they are depicted in Fig.2. The generation rates, $G_{\rm CVCC}(n,p)$ and  $G_{\rm VVCV}(p,n)$, also satisfy $G_{\rm CVCC}(n,p) = G_{\rm VVCV}(p,n)$. In thermal equilibrium, the total generation and recombination rates must be equal, i.e. $G_{\rm CVCC} + G_{\rm VVCV} = R_{\rm CCCV} + R_{\rm CVVV}$. 

Given the symmetry between the various processes, it is enough to consider in detail only one process. In the discussion that follows we will concentrate only on the CCCV Auger recombination process. In the CCCV process, energy conservation implies,
\begin{eqnarray}
\hbar v |{\bf k}_{1}| + \hbar v |{\bf k}_{2}| & = & \hbar v |{\bf k}_{1}+{\bf Q}| - \hbar v |{\bf k}_{2}-{\bf Q}| \nonumber \\
|{\bf k}_{1}| + |{\bf k}_{2}| & = & |{\bf k}_{1}+{\bf Q}| - |{\bf k}_{2}-{\bf Q}| 
\end{eqnarray}
The available phase space for Auger recombination can be understood as follows. For any three vectors ${\bf k}_{1}$, ${\bf k}_{2}$, and  ${\bf Q}$ one has the identity,
\begin{equation}
|{\bf k}_{1}+{\bf Q}| - |{\bf k}_{2}-{\bf Q}| \le |{\bf k}_{1}+{\bf k}_{2}| \le |{\bf k}_{1}| + |{\bf k}_{2}|
\end{equation}
The energy conservation condition requires both the inequalities above to be equalities. The inequality on the right would be an equality if (and only if) the vectors ${\bf k}_{1}$ and ${\bf k}_{2}$ point in the same direction. The inequality on the left would be an equality if (and only if) the vectors ${\bf k}_{1}+{\bf Q}$ and ${\bf k}_{2}-{\bf Q}$ point in the opposite direction. If ${\bf k}_{1}$ and ${\bf k}_{2}$ point in the same direction, then ${\bf k}_{1}+{\bf Q}$ and ${\bf k}_{2}-{\bf Q}$ will point in the opposite direction if (and only if) ${\bf k}_{1}$ and ${\bf Q}$ also point in the same direction and $|{\bf Q}|>|{\bf k}_{2}|$. Energy conservation therefore requires that the vectors ${\bf k}_{1}$, ${\bf k}_{2}$, and  ${\bf Q}$ all lie on the same line. This requirement also holds for all the Coulomb scattering processes depicted in Fig.1 and Fig.2. 

The Bloch functions for the conduction ($s=+1$) and valence ($s=-1$) band electrons in Graphene can be written as~\cite{dressel},
\begin{equation}
\psi_{s,{\bf k}}(r)=\frac{e^{i{\bf k}.{\bf r}}}{\sqrt{N}} \, \, u_{s,{\bf k}}(r)
\end{equation}
Here, $N$ is the total number of unit cells in the crystal. The periodic part $u_{s,{\bf k}}(r)$ of the Bloch function has the following overlap integral~\cite{dressel},
\begin{equation}
 |<u_{s',{\bf k'}}|u_{s,{\bf k}}>|^{2} = \frac{1}{2} \, \left[ 1 + s s' \, \cos(\theta_{{\bf k'},{\bf k}}) \right]
\end{equation}
where $\theta_{{\bf k'},{\bf k}}$ is the angle between the vectors ${\bf k'}$ and ${\bf k}$. We assume that the occupation statistics of electrons in conduction and valence bands are given by the Fermi-Dirac distribution functions $f_{s}({\bf k})$,
\begin{equation}
f_{s}({\bf k}) = \frac{1}{1 + e^{(E_{s}({\bf k}) - E_{fs})/KT}}
\end{equation}
$E_{fs}$ are the Fermi levels. We assume different Fermi levels for conduction and valence electrons to allow for non-equilibrium electron-hole populations, as is the case in a forward biased pn-junction diode~\cite{pierret}. The electron and hole densities are given as follows,
\begin{equation}
n = 4  \int \frac{d^{2}{\bf k}}{(2 \pi)^{2}} \, f_{+1}({\bf k})
\end{equation}
\begin{equation}
p = 4  \int \frac{d^{2}{\bf k}}{(2 \pi)^{2}} \, \left[1 - f_{-1}({\bf k}) \right]
\end{equation}
The factor of 4 in the front accounts for spin degeneracy and the two valleys at $K$ and $K'$. 

The electron-hole recombination rate $R_{CCCV}(n,p)$ (units: cm$^{-2}$-sec$^{-1}$) due to Auger scattering can be written as~\cite{many},
\begin{eqnarray}
R_{\rm CCCV}(n,p) & = & 2 \, \left( \frac{2 \pi}{\hbar} \right) \, \int \frac{d^{2}{\bf k}_{1}}{(2 \pi)^{2}} \,  \int \frac{d^{2}{\bf k}_{2}}{(2 \pi)^{2}} \,  \int \frac{d^{2}{\bf Q}}{(2 \pi)^{2}} \, |M({\bf k}_{1}, {\bf k}_{2}, {\bf Q})|^{2} \, \nonumber \\
& & \left[ 1 - f_{-1}({\bf k}_{2}-{\bf Q}) \right] \,  \left[ 1 - f_{+1}({\bf k}_{1}+{\bf Q}) \right] \,  f_{+1}({\bf k}_{1}) \,  f_{+1}({\bf k}_{2}) \, \nonumber \\
& & \delta(\hbar v |{\bf k}_{1}| + \hbar v |{\bf k}_{2}| - \hbar v |{\bf k}_{1}+{\bf Q}| + \hbar v |{\bf k}_{2}-{\bf Q}|) \label{w1}
\end{eqnarray}
The factor of two in the front comes from the two valleys at $K$ and $K'$. The scattering matrix element $M({\bf k}_{1}, {\bf k}_{2}, {\bf Q})$ includes both direct and exchange processes, and can be written as,
\begin{equation}
|M({\bf k}_{1}, {\bf k}_{2}, {\bf Q})|^{2} = |M_{d}({\bf k}_{1}, {\bf k}_{2}, {\bf Q})|^{2} + |M_{e}({\bf k}_{1}, {\bf k}_{2}, {\bf Q})|^{2} + |M_{d}({\bf k}_{1}, {\bf k}_{2}, {\bf Q}) - M_{e}({\bf k}_{1}, {\bf k}_{2}, {\bf Q})|^{2}
\end{equation}
Assuming statically screened Coulomb interaction, the matrix elements, $M_{d}({\bf k}_{1}, {\bf k}_{2}, {\bf Q})$ and $M_{e}({\bf k}_{1}, {\bf k}_{2}, {\bf Q})$, for the direct and exchange scattering processes, respectively, are as follows,
\begin{equation}
M_{d}({\bf k}_{1}, {\bf k}_{2}, {\bf Q}) =  \frac{e^{2}}{2 \, \epsilon_{\infty} \, \left( |{\bf Q}| + Q_{TF} \right)} \, <u_{+1,{\bf k}_{1}+{\bf Q}}|u_{+1,{\bf k}_{1}}> \, <u_{-1,{\bf k}_{2}-{\bf Q}}|u_{+1,{\bf k}_{2}}> \label{me} 
\end{equation}
\begin{equation}
M_{e}({\bf k}_{1}, {\bf k}_{2}, {\bf Q}) = \frac{e^{2}}{2 \, \epsilon_{\infty} \, \left( |{\bf Q}+{\bf k}_{1}-{\bf k}_{2}| + Q_{TF} \right)} \, <u_{+1,{\bf k}_{1}+{\bf Q}}|u_{+1,{\bf k}_{2}}> \, <u_{-1,{\bf k}_{2}-{\bf Q}}|u_{+1,{\bf k}_{1}}> \label{md} 
\end{equation}
Here, $e$ is the electron charge, $Q_{TF}$ is the Thomas-Fermi wavevector~\cite{darma}, and $\epsilon_{\infty}$ is the average of the dielectric constants of the media on both sides of the Graphene layer. The relative directions for the vectors ${\bf k}_{1}$, ${\bf k}_{2}$, and  ${\bf Q}$ allowed by energy conservation results in the values of all the overlap integrals in Equations (\ref{me}) and (\ref{md}) to equal unity. Assuming screening by both electrons and holes, the expression for the Thomas-Fermi wavevector in Graphene is~\cite{darma},
\begin{equation}
Q_{TF} = \frac{e^{2} \, KT}{\pi \, \epsilon_{\infty} \hbar^{2} \, v^{2}} \log{\left[ \left( e^{E_{f+1}/KT} + 1 \right) \left( e^{-E_{f-1}/KT} + 1 \right) \right]}
\end{equation}
where, $E_{f+1}$ and $E_{f-1}$ are the Fermi levels for the conduction and valence electrons, respectively. After integrating out the delta function, the six-dimensional integral in Equation (\ref{w1}) can be reduced to the following three-dimensional integral,
\begin{eqnarray}
R_{\rm CCCV}(n,p) & = & \frac{1}{\hbar^{2} v} \, \int_{0}^{\infty} \frac{d k_{1}}{2 \pi} \,  \int_{0}^{\infty} \frac{d k_{2}}{2 \pi } \,  \int_{k_{2}}^{\infty} \frac{d Q}{2 \pi} \, |M(k_{1}, k_{2}, Q)|^{2} \,  \sqrt{ (k_{1}+Q) \, (Q-k_{2}) \, k_{1} \, k_{2} } \nonumber \\
& & \left[ 1 - f_{-1}(Q-k_{2}) \right] \,  \left[ 1 - f_{+1}(k_{1}+Q) \right] \,  f_{+1}(k_{1}) \,  f_{+1}(k_{2}) \label{w2}
\end{eqnarray}
The above equation is the main result of this paper. The total Auger recombination rate $R(n,p)$ is the sum of the rates of the CCCV and CVVV processes, 
\begin{equation}
R(n,p) = R_{\rm CCCV}(n,p) + R_{\rm CVVV}(n,p)
\end{equation}
The average electron-hole recombination time $\tau_{r}$ due to Coulomb scattering is defined as,
\begin{equation}
\frac{1}{\tau_{r}} = \frac{R(n,p)}{{\rm min}(n,p)} \label{eqtau}
\end{equation} 
where the smaller carrier density appears in the denominator on the right hand side. $\tau_{r}$ can also be interpreted as the minority carrier lifetime in situations where the electron and hole densities are very different (see the discussion below). Using the result in Equation (\ref{w2}), the generation rate $G_{\rm CVCC}(n,p$) due to CVCC process can be written as,
\begin{eqnarray}
G_{\rm CVCC}(n,p) & = & \frac{1}{\hbar^{2} v} \, \int_{0}^{\infty} \frac{d k_{1}}{2 \pi} \,  \int_{0}^{\infty} \frac{d k_{2}}{2 \pi } \,  \int_{k_{2}}^{\infty} \frac{d Q}{2 \pi} \, |M(k_{1}, k_{2}, Q)|^{2} \,  \sqrt{ (k_{1}+Q) \, (Q-k_{2}) \, k_{1} \, k_{2} } \nonumber \\
& & \left[ 1 - f_{+1}(k_{1}) \right] \,  \left[ 1 - f_{+1}(k_{2)} \right] \,  f_{-1}(Q-k_{2}) \,  f_{+1}(Q+k_{1}) \label{w3}
\end{eqnarray}
The total generation rate $G(n,p)$ is the sum of the rates of the CVCC and VVCV processes, 
\begin{equation}
G(n,p) = G_{\rm CVCC}(n,p) + G_{\rm VVCV}(n,p)
\end{equation}
The average electron-hole generation time $\tau_{g}$ due to Coulomb scattering is defined as,
\begin{equation}
\frac{1}{\tau_{g}} = \frac{G(n,p)}{{\rm min}(n,p)} \label{eqtau2}
\end{equation} 

\begin{figure}[tbp]
  \begin{center}
   \epsfig{file=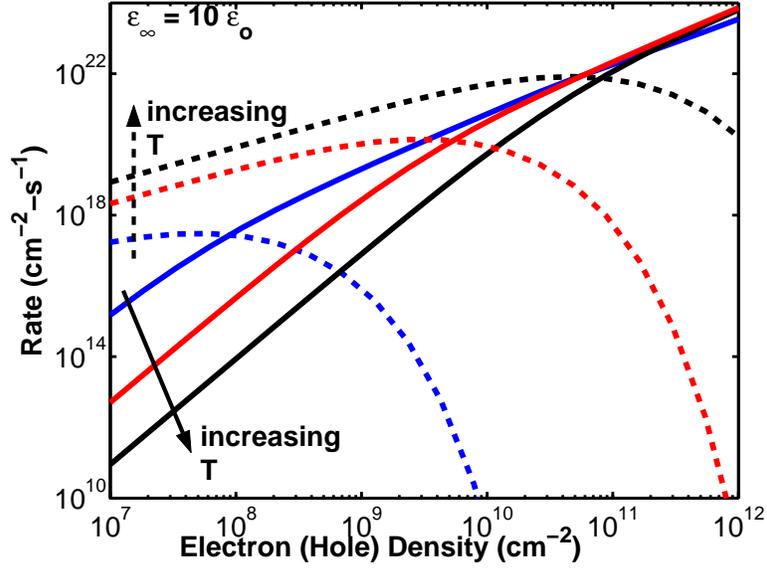,angle=0,width=4.0 in}    
    \caption{Electron-hole generation (dashed) and recombination (solid) rates are plotted as a function of the electron and hole densities (assumed to be equal) for different temperatures. The different curves correspond to temperatures T=10K, 77K, and 300K. The assumed values of $v$ and $\epsilon_{\infty}$ are $10^{8}$ cm/s and $10 \epsilon_{o}$, respectively.} 
    \label{Fig3}
  \end{center}
\end{figure}
\begin{figure}[tbp]
  \begin{center}
   \epsfig{file=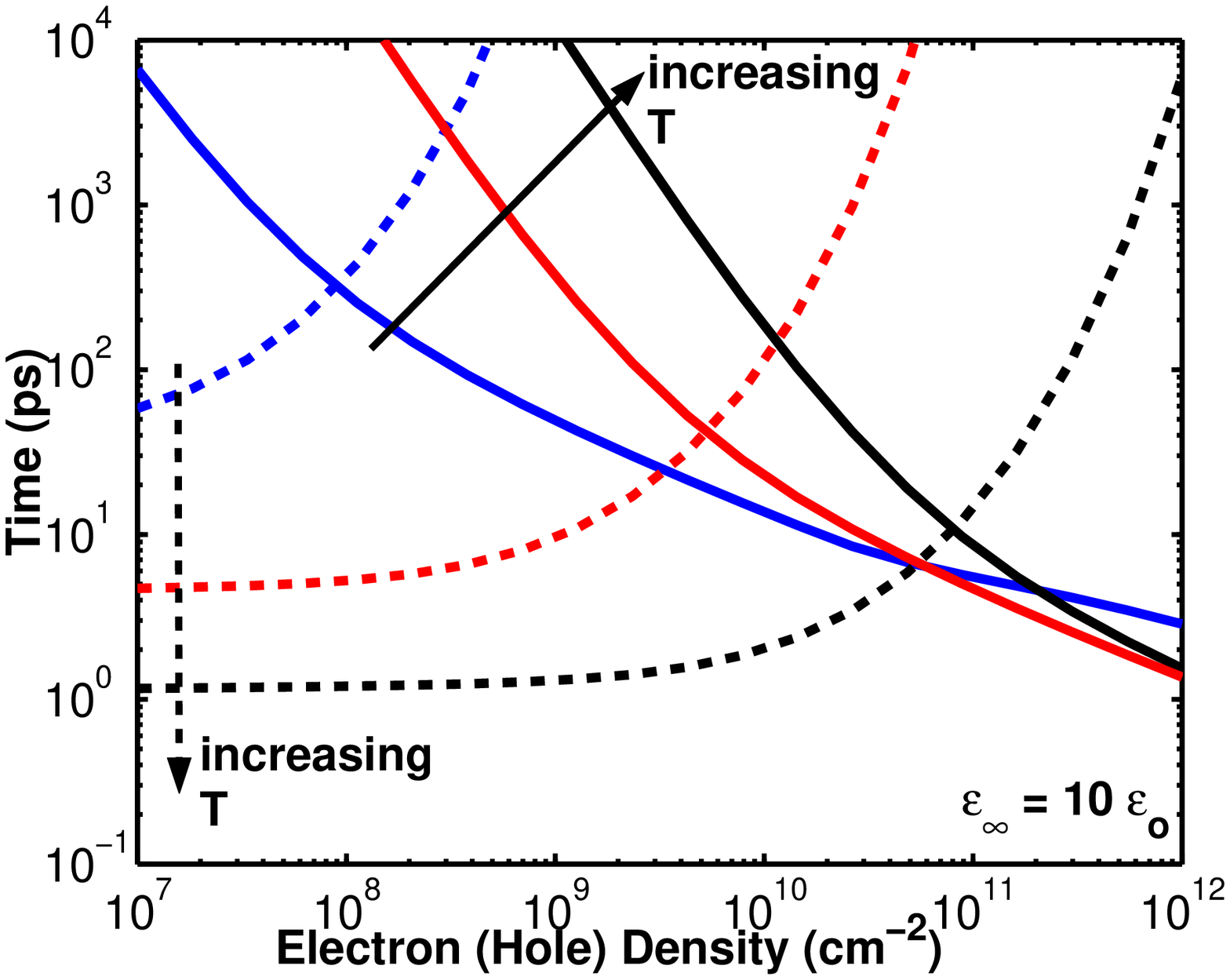,angle=0,width=4.0 in}    
    \caption{Electron-hole generation (dashed) and recombination (solid) times are plotted as a function of the electron and hole densities (assumed to be equal) for different temperatures. The different curves correspond to temperatures T=10K, 77K, and 300K. The assumed values of $v$ and $\epsilon_{\infty}$ are $10^{8}$ cm/s and $10 \epsilon_{o}$, respectively.}
    \label{Fig4}
  \end{center}
\end{figure}
\begin{figure}[tbp]
  \begin{center}
   \epsfig{file=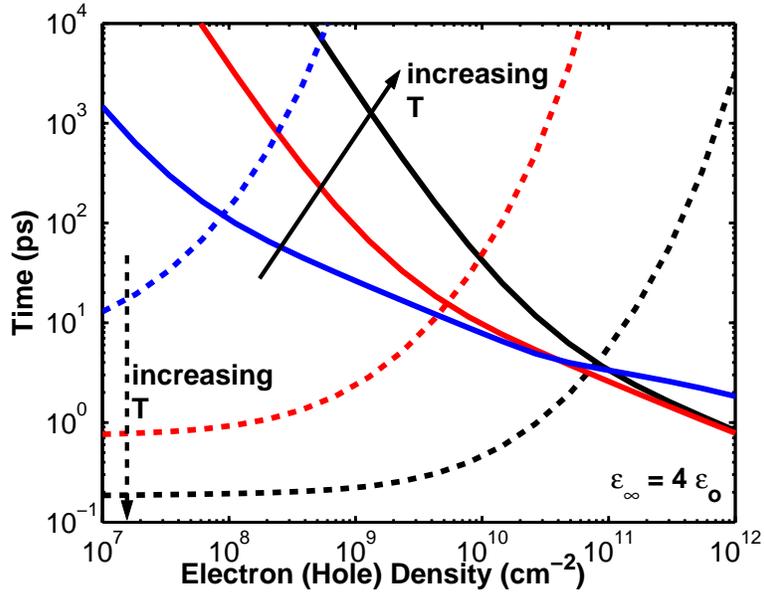,angle=0,width=4.0 in}    
    \caption{Electron-hole generation (dashed) and recombination (solid) times are plotted as a function of the electron and hole densities (assumed to be equal) for different temperatures. The different curves correspond to temperatures T=10K, 77K, and 300K. The assumed values of $v$ and $\epsilon_{\infty}$ are $10^{8}$ cm/s and $4 \epsilon_{o}$, respectively.}
    \label{Fig5}
  \end{center}
\end{figure}

\section{Results}
Fig.3 shows the total generation and recombination rates plotted as a function of the electron and hole densities, which are assumed to be equal, for different temperatures (T=10K, 77K, and 300K). The value of $\epsilon_{\infty}$ used in simulations is $10 \epsilon_{o}$ assuming Aluminum-oxide on both sides of the Graphene layer~\cite{marcus}. Fig.4 shows the corresponding generation and recombination times. For any temperature, the generation and recombination rate curves cross at the point where the electron and hole densities have their thermal equilibrium values. Just like in two dimensional semiconductor quantum wells, the temperature dependence of Coulomb scattering is a sensitive function of the electron and hole densities as well as temperature~\cite{paul}. At higher temperatures the probability of finding energetic electrons and holes is larger and therefore the generation rate increases with temperature. For small electron-hole densities, increase in temperature spreads the carrier distributions to higher energies where Auger recombination is less efficient and therefore the recombination rate decreases. But for large electron-hole densities, the electrons and holes near the band edges can recombine only if the final scattering states are unoccupied. An increase in temperature generates more unoccupied states. As a result of the above two factors, the recombination rates at large electron-hole densities are less sensitive to temperature. 

Fig.4 shows that for electron-hole densities smaller than $10^{12}$ cm$^{-2}$ the recombination time is longer than 1 ps at all temperatures, and for electron-hole densities smaller than  $10^{11}$ cm$^{-2}$ the recombination time is longer than 5 ps at all temperatures. Fig.5 shows the generation and recombination times for $\epsilon_{\infty}=4 \epsilon_{o}$ assuming Silicon-dioxide on both sides of the Graphene layer. Clearly, the Coulomb scattering rates depend on the dielectric surrounding the Graphene layer and are enhanced for a smaller dielectric constant medium. The effect of the surrounding dielectric on the generation and recombination times is more pronounced at small electron-hole densities when carrier screening of the Coulomb interaction is less effective. 

\begin{figure}[tbp]
  \begin{center}
   \epsfig{file=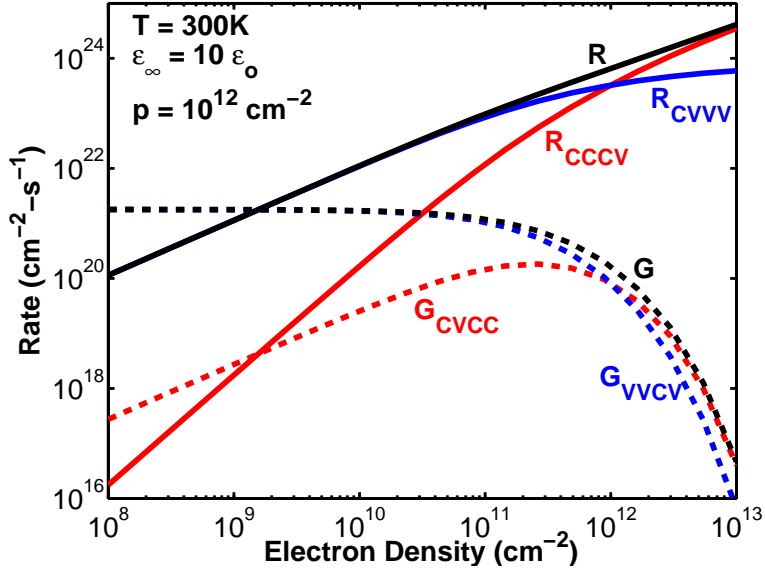,angle=0,width=4.0 in}    
    \caption{Electron-hole generation (dashed) and recombination (solid) rates are plotted for different electron densities. The hole density is fixed and equals $10^{12}$ cm$^{-2}$. T=300K. The contributions $G_{\rm CVCC}$ and $G_{\rm VVCV}$ to the total generation rate $G$, and the contributions $R_{\rm CCCV}$ and $R_{\rm CVVV}$ to the total recombination rate $R$, are also plotted.}
    \label{Fig6}
  \end{center}
\end{figure}
\begin{figure}[tbp]
  \begin{center}
   \epsfig{file=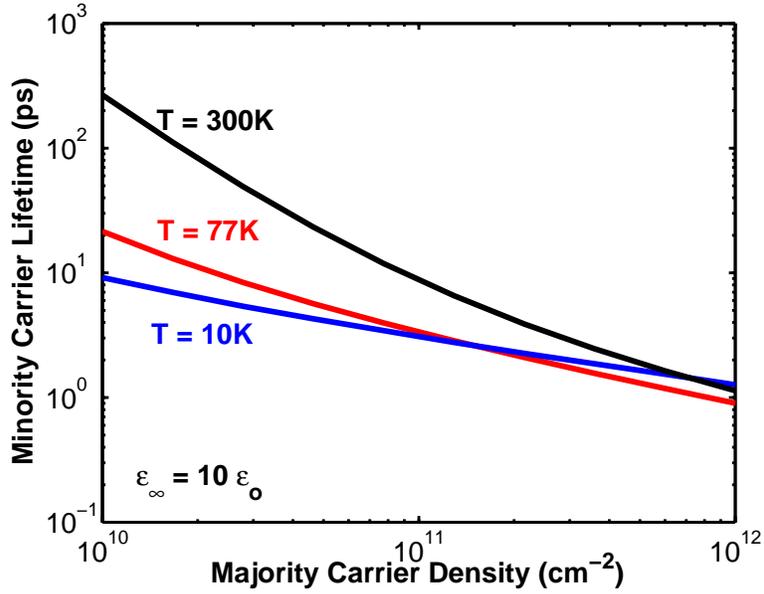,angle=0,width=4.0 in}    
    \caption{Minority carrier (electron) lifetimes is plotted as a function of the majority (hole) density for different temperatures (T=10K, 77K, and 300K). The assumed values of $v$ and $\epsilon_{\infty}$ are $10^{8}$ cm/s and $10 \epsilon_{o}$, respectively.}
    \label{Fig7}
  \end{center}
\end{figure}

Fig.6 shows the individual contributions, $R_{\rm CCCV}$ and $R_{\rm CVVV}$ of the CCCV and CVVV processes, respectively, to the total Auger recombination rate $R$, and the contributions $G_{\rm CVCC}$ and $G_{\rm VVCV}$ of the CVCC and VVCV processes, respectively, to the total generation rate $G$. The electron density is varied and the hole density is fixed at $10^{12}$ cm$^{-2}$. T=300K and $\epsilon_{\infty}=10 \epsilon_{o}$. The thermal equilibrium value of the electron density, corresponding to the hole density of $10^{12}$ cm$^{-2}$, is $1.5\times 10^{9}$ cm$^{-2}$. For electron densities much smaller than the hole density the CVVV process dominates recombination, and the recombination rate varies approximately linearly with the electron density. For electron densities much larger than the hole density the CCCV process dominates recombination. When electron and hole densities are equal then, as explained earlier, $R_{\rm CCCV}=R_{\rm CVVV}$. For electron densities much smaller than the hole density the VVCV process dominates generation and the generation rate is almost independent of the electron density. For electron densities much larger than the hole density the CVCC process dominates generation. For equal electron and hole densities, $G_{\rm CVCC}=G_{\rm VVCV}$. 

For device applications, it is also interesting to look at the minority carrier generation-recombination rates in situations where the electron (or the hole) density is much smaller than the hole (or the electron) density. This is the case, for example, in a forward biased pn-junction~\cite{pierret}. Fig.6 and the discussion above shows that the minority carrier generation and recombination rates can be written approximately as follows,
\begin{equation}
R-G = \frac{n-n_{o}}{\tau_{r}}
\end{equation}
where, we have assumed that the electrons are the minority carriers and holes are the majority carriers. $n_{o}$ is the thermal equilibrium electron density and $\tau_{r}$ is the minority carrier (electron) lifetime. The minority carrier lifetime is independent of the minority carrier density but depends on the majority carrier density and the temperature. From Fig.6, the minority carrier lifetime for a majority carrier density of $10^{12}$ cm$^{-2}$ is approximately 1.1 ps at T=300K. Fig.7 shows the minority carrier (electron) lifetime as a function of the majority carrier (hole) density for a different temperatures (T=10K, 77K, and 300K). $\epsilon_{\infty}=10 \epsilon_{o}$. 

In conclusion, we have calculated electron-hole generation and recombination rates due to Coulomb scattering in Graphene. Our results show that electron-hole recombination times in Graphene can be much longer than 1 ps at all temperatures for electron-hole densities smaller than $10^{12}$ cm$^{-2}$. The author would like to acknowledge helpful discussions with Edwin Kan and Sandip Tiwari.

\end{document}